\documentclass[twoside,fleqn]{article}
\usepackage{espcrc2,texdraw,epsf,feynmp}
\newcommand{\bfx}{\mathbf{x}}
\newcommand{\bfp}{\mathbf{p}}
\title{The second moment of the pion's distribution amplitude
\thanks{talk presented by Luigi Del Debbio}}
\author{L. Del Debbio, M. Di Pierro, A. Dougall and C. Sachrajda
(UKQCD Collaboration) \vskip 5mm
Dept. of Physics and Astronomy, Univ. of Southampton,
Southampton SO17 1BJ, UK \\
}
\begin{document}

\begin{abstract}
We present preliminary results for the second moment of the pion's
distribution amplitude. The lattice formulation and the
phenomenological implications are briefly reviewed, with special
emphasis on some subtleties that arise when the Lorentz group is
replaced by the hypercubic group. Having analysed more than half of
the available configurations, the result obtained is $\langle \xi^2
\rangle_L = 0.06\pm 0.02$.
\end{abstract}

\maketitle

\section{INTRODUCTION}
The distribution amplitude $\phi(x_1,x_2)$ gives the probability for
finding two collinear partons with fractions $x_1$ and $x_2$ of the
meson's momentum~\cite{chernyak84}. It plays an important role e.g. in
exclusive hard scattering processes and in non-leptonic decays of
heavy mesons, since it allows one to separate a hard scattering
amplitude, computed in perturbation theory, from the effect of the
soft Fourier modes, whose dynamics is non-perturbative.

\begin{figure}[htb]
\begin{center}

\begin{fmffile}{pion}
\begin{fmfgraph*}(150,75)

\fmfleft{ip}
\fmfright{op}
\fmftop{oph}

\fmfrpolyn{tension=0.25,filled=empty,label=$\phi$,smooth}{G}{3}
\fmfrpolyn{tension=0.25,filled=empty,label=$\phi$,smooth}{H}{3}

\fmf{photon, tension=0}{oph,v2}

\fmf{double,label=$p$}{ip,G1}
\fmf{vanilla,tension=0.4,label=\small{$x_1$}}{G2,v2}
\fmf{vanilla,tension=0.4}{v2,v3}
\fmf{vanilla,tension=0.4,label=\small{$y_1$},label.side=left}{v3,H3}
\fmf{vanilla,tension=0.4,label=\small{$x_2$},label.side=right}{G3,v6}
\fmf{vanilla,tension=0.4}{v6,v4}
\fmf{vanilla,tension=0.4,label=\small{$y_2$},label.side=right}{v4,H2}
\fmf{double,label=$p^\prime$,label.side=left}{H1,op}
\fmffreeze
\fmf{gluon}{v3,v4}

\end{fmfgraph*}
\end{fmffile}

\end{center}
\vspace{-.6in}
\caption{Photon-pion interaction, factorised into a hard
(perturbative) scattering amplitude and a non-perturbative part
described by the pion distribution amplitude.}
\label{fig:def}
\end{figure}
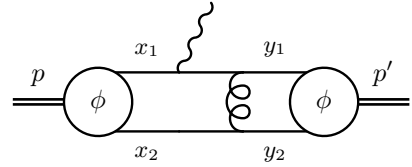

For example, the relevant diagram for the pion electromagnetic form
factor is shown in Fig.~\ref{fig:def}. The form factor is defined as:
\begin{equation}
\langle \pi(p^\prime) | V_\mu | \pi(p) \rangle = F(q^2) (p+p^\prime)_\mu
\end{equation}
and can be written in terms of the distribution amplitude: 
\begin{eqnarray}
\lefteqn{F(Q^2) =} \nonumber \\ 
&& \int [{\mathrm d}x][{\mathrm d}y]\: \phi^\dagger(x,Q^2) 
T_H(x,y;Q^2) \phi(y,Q^2)
\end{eqnarray}
where $[{\mathrm d}x]={\mathrm d}x_1 {\mathrm d}x_2
\delta(x_1+x_2-1)$, and $T_H$ is the perturbative amplitude computed
at the quark level.

The amplitude $\phi$ is given by the non-local Fourier transform of
the matrix element of the fermionic bilinear~\cite{brodsky80}:
\begin{equation}
\phi^{ab}_{\alpha\beta}(x,Q^2) \sim
FT \left[ \langle 0 | T \psi^{a}_{\alpha}(z_1) \bar\psi^{b}_{\beta}(z_2) | \pi
\rangle \right]
\label{eq:phidef}
\end{equation}
A light-cone expansion relates $\phi$ to fermion bilinears with
covariant derivatives. In particular the $m$-th moment,
\begin{equation}
\langle \xi^m \rangle \equiv \int {\mathrm d}\xi \: \xi^m \phi(\xi,Q^2),
\end{equation}
can be extracted from  the matrix element of the lowest-twist operator
appearing in the OPE of Eq.~\ref{eq:phidef}:
\begin{equation}
\label{eq:def1}
\langle 0 | O_{\mu_0 \ldots \mu_m}(0) | \pi(p)\rangle =
        f_\pi\: p_{\mu_0} \ldots p_{\mu_m} \: \langle \xi^m \rangle + 
        \cdots  
\end{equation}
where
\begin{equation}
\label{eq:def2}
O_{\mu_0 \ldots \mu_m}(0) = (-i)^n \bar\psi \: \gamma_{\mu_0} \gamma_5
        \:
        \stackrel{\leftrightarrow}{D}_{\mu_1}\ldots
        \stackrel{\leftrightarrow}{D}_{\mu_m}\psi
\end{equation}
and the ellipses indicate terms that are proportional to $p^2
g_{\mu_i \mu_j}$.

The distribution amplitude has been studied in recent years using sum
rules (see results quoted in~\cite{chernyak84}) and lattice
simulations~\cite{martinelli87,gupta91}. The accuracy of the lattice
determinations so far has not been good enough to compare precisely
with sum rules predictions. The aim of the current study is to use
modern lattice technology to improve the precision of the result.

\section{LATTICE DETAILS}
The following UKQCD quenched configurations have been used to compute the
relevant matrix elements. 
\begin{itemize}
\item $\beta=6.2$, $a^{-1}=2.64\pm0.10~\mbox{GeV}$, $24^3\times 48$ lattice;
\item SW action, with $c_{\mathrm{sw}}=1.61$;
\item the three $\kappa$ values used in our simulation (0.13460,
0.13510, 0.13530) cor\-res\-pond to physical light pseudoscalar masses
ranging from 350 to 850~MeV.
\item 180 configurations are available, the results presented here are
based on the analysis of a subset of 129;
\item non-improved local operators have been used, without fuzzing of
the light quarks.
\end{itemize}

\section{POWER-LIKE DIVERGENCIES}

In this work, we concentrate on the second moment of the distribution
amplitude, i.e. $m=2$ in Eq.~\ref{eq:def1} and~\ref{eq:def2}
above. The choice of the Lorentz indices for $O_{\mu_0\mu_1\mu_2}$ is
crucial in order to avoid mixing with lower-dimensional operators. We
choose $\mu_0$ to be the time-direction. It is convenient, but not
necessary, to avoid time-derivatives and hence we do not symmetrise
$\mu_0$ with $\mu_1,\mu_2$. Using the notation of~\cite{mandula80}
to classify the irreducible representations of the hypercubic group,
we obtain:
\begin{itemize}
\item $O_{\mu[\nu\sigma]}$, with
  $\mu\!\neq\!\nu\!\neq\!\sigma\!\neq\!\mu$ and symmetrised over $\nu$
  and $\sigma$, transforms like a
  $\overline{(\frac{1}{2},\frac{1}{2})} \oplus \mathbf{8}$ reducible
  representation, and therefore does not mix with lower dimensional
  operators. However the $\mathbf{8}$ irrep leads to a term
  proportional to $p_\mu \left[ \frac{p_\nu^2 + p_\sigma^2}{2} -
  p_\tau^2\right]$, where $\tau$ is the fourth available index, which
  needs to be subtracted in order to have an operator proportional to
  $p_\mu p_\nu p_\sigma$, as in the continuum limit. This subtraction
  is performed in the definition of $R_1$ below. Note that two spatial
  components of $p$ need to be non-zero to get a non-vanishing signal.
\item $O_{\mu\nu\nu}$, with $\mu\!\neq\!\nu$ transforms like a
  $(\frac{1}{2},\frac{1}{2}) \oplus \mathbf{8}$ representation;
  however the subtracted operator:
\[
 O_{411}^\prime=\left(O_{411}-\frac{O_{422}+O_{433}}{2}\right)
\]
simply transforms like $\mathbf{8}$. In this case there is no mixing,
and a non-zero signal is obtained with just one non-vanishing
component of $p=(1,0,0)$.
\end{itemize}

\section{PRELIMINARY RESULTS}
The relevant matrix elements are extracted from suitably defined two-point
functions. For a generic operator $Q$:
\begin{eqnarray}
C_2^{Q}(t) &=&
        \sum_\bfx e^{i \bfp \bfx} \langle Q(\bfx,t) \: \bar\psi \gamma_5
        \psi(0) \rangle 
        \nonumber \\
        &\stackrel{t\rightarrow\infty}{\rightarrow}&
        \frac{Z}{2E} \: \langle 0 | Q(0) | \pi(p)\rangle \: e^{-Et}
\end{eqnarray}
where $Z=\langle \pi(p)| \bar\psi\gamma_5\psi(0) |0\rangle$.

\noindent
For large time separations:
\begin{eqnarray}
R_1 &=& \left.\frac{C_2^{O}(t)}{C_2^{A}(t)}\right|_{\mathbf{p}=(1,1,0)} -
      \left.\frac{C_2^{O}(t)}{C_2^{A}(t)}\right|_{\mathbf{p}=(1,0,0)}
      \nonumber \\ 
    &=& p_1\, p_2 \langle \: \xi^2 \rangle
\nonumber \\
R_2 &=& {C_2^{O^\prime}(t)}/{C_2^{A}(t)} \nonumber \\
    &=& p_1^2 \: \langle \xi^2 \rangle
\nonumber 
\end{eqnarray}

\begin{figure}[ht]
\centerline{\setlength\epsfxsize{0.9\hsize} \setlength\epsfysize{0.65\hsize}
\epsfbox{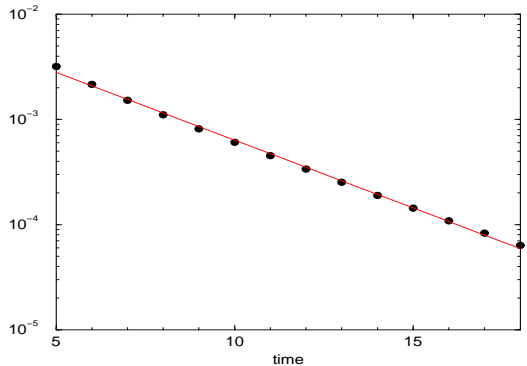}}
\vspace{-.3in}
\caption{Pion propagator from 129 configurations at $\kappa=0.13460$}
\label{fig:pionprop}
\end{figure}

Only the results for the heavier $\kappa$ are presented here. The pion
propagator is shown in Fig.~\ref{fig:pionprop}. The effective mass and
single-exponential fits give consistent results for $6<t<15$. Since
the relative error on the signal of interest for the second moment
becomes large for $t>15$, the fitting ranges chosen in this work are
shorter than the ones usually used for light spectroscopy. Despite
this discrepancy, our value for $f_\pi$ agrees with other UKQCD
determinations from the same set of configurations within $2\sigma$.

\begin{figure}[ht]
\centerline{\setlength\epsfxsize{0.9\hsize} \setlength\epsfysize{0.65\hsize}
\epsfbox{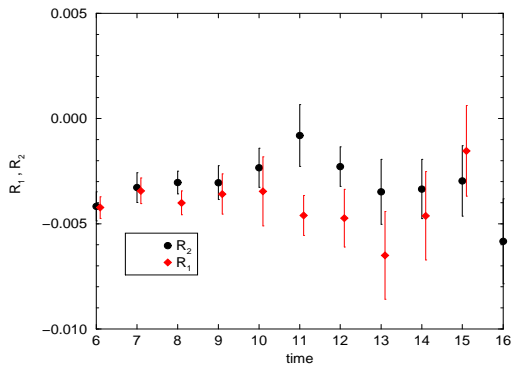}}
\vspace{-.3in}
\caption{Ratios of two-point functions $R_1$ and $R_2$ for
$\kappa=0.13460$. The results for $R_1$ and $R_2$ are obtained from 55
and 129 configurations respectively.} 
\label{fig:R1}
\end{figure}

The time dependence of $R_2$, obtained from 129 configurations, is
given in Fig.~\ref{fig:R1}. We have currently analysed $R_1$ on 55
configurations only. In order to compare the two determinations, the
results for $R_1$ are also displayed in the same figure. The agreement
between the two sets of results is good. At this preliminary stage
$R_2$ alone is used to extract our final number.

\section{CONCLUSIONS}

Fitting the points in Fig.~\ref{fig:R1} to a constant yields 
\[
\langle \xi^2 \rangle_L = 0.06 \pm 0.02
\]
for the lattice value of the second moment of the pion distribution
amplitude. The analysis of the complete set of available
configurations for both $R_1$ and $R_2$, together with a full
discussion of the systematic errors, will be presented in a
forthcoming publication~\cite{realpi}.  The statistical error in our
result will of course decrease as the number of configurations is
increased. The multiplicative renormalisation of the operator has yet
to be calculated, either perturbatively or non-perturbatively.

Our preliminary result for the value of $\langle \xi^2 \rangle$ suggests
that it is lower than the value predicted by sum rules. Such a result
could be explained for example by a more peaked distribution for $\xi$
around the origin.

Computations of the distribution amplitudes of the $\rho$ meson and
the proton are currently under way. Our analysis can also be extended
to the case of non-degenerate quarks and so a study of the
distribution amplitudes of the $K$ and $B$ mesons is feasible.


\begin{thebibliography}{9}
\bibitem{chernyak84} for a review and extensive references, see
V.L.~Chernyak, I.R.~Zhitnitsky, Phys. Rep. {\bf 112} (1984) 173
\bibitem{brodsky80} S.~Brodsky et al., Phys. Lett. {\bf 91 B} (1980) 239
\bibitem{martinelli87}  G.~Martinelli and C.~Sachrajda,
Phys. Lett. {\bf B 190} (1987) 151 
\bibitem{gupta91} T.~Daniel, R.~Gupta, D.~Richards, Phys. Rev. {\bf D
43} 3715 (1991) 
\bibitem{mandula80} J.~Mandula, G.~Zweig, J.~Govaerts,
Nucl. Phys. {\bf B 228} (1980) 91 
\bibitem{realpi} L.~Del~Debbio, M.~Di~Pierro, A.~Dougall,
C.T.~Sachrajda, in preparation
\end{thebibliography}
\end{document}